\documentclass[12pt,a4paper]{article}

\usepackage{a4}
\usepackage{epsfig}
\usepackage{bbm}
\usepackage{mathptm}

\def\R{\mathbbm{R}}
\def\Z{\mathbbm{Z}}

\begin{document}

\title{\Huge String Theory and Beyond}
\author{M. Rudolph \\
	Institut f\"ur Theoretische Physik, Univ. Leipzig \\
	Augustusplatz 10/11, 04109 Leipzig, Germany \\
	e-mail: michael.rudolph@itp.uni-leipzig.de}	       
\maketitle

\begin{abstract}
This is the written version of a short talk given at the University of Leipzig 
in December 1998. It reviews some general aspects of string theory from the 
viewpoint of the search for an unifying theory. Here, special emphasis lies 
on the motivation to consider string theory not only as the leading candidate 
for the unification of gravity and the other fundamental forces of nature, 
but also as a possible step towards a new understanding of nature and its 
description within the framework of physical models. Without going into 
details, some recent developments, including duality symmetries and the 
appearance of $M$--theory, are reviewed.
\end{abstract}

\section{Introduction or {\it Why Strings?}}
\label{Introduction}

In the history of science there are only a few examples of theories and 
thoughts that influenced the way of physical research, mathematics and even 
natural philosophy to such a lasting extend like the theory of strings. But 
at the same time it splits the physical community into two nearly disjunct 
groups: one group whose representatives believe that string theory will provide 
a right way to the solution of the unifying problem of (theoretical) physics,
and a second group whose members are convinced that string theory is nothing 
but another dead end in the long search for answers in physics.

As a young physicist just entering the active research in this field I do not 
want to choose one of this groups. In contrast I'm guided by a more diplomatic 
viewpoint of one of my former lecturers, a mathematician, who said that 
{\it ``string theory has brought many new and fascinating ideas to life and 
stimulated many new researches, so it cannot be completely wrong.''} Thus, 
unanswered the question if string theory is the last answer to the open 
problems in theoretical physics today, it appears to be at least a small part 
of the truth each scientist should strive for. With this in mind it surely 
cannot be a failure to deal with the fascinating topics arising from it.

In what follows I want to present some arguments which justify this viewpoint. 
Of course it is not possible to give a detailed introduction into string theory,
even at a very coarse level, within this short talk. Therefore I want to 
restrict to some general aspects (maybe of more philosophical nature) of 
string theory and its recent status. Indeed, string theory demands such a 
deep change of the basic principles of our physical understanding, that a 
more general viewpoint is necessary to experience its beauty and richness.

In that sense, let me start in a somewhat unconventional way compared to 
common introductionary talks to strings, namely the question about {\it how 
a grand unifying theory of physic --- or the Theory of Everything (TOE) 
\cite{Wei92a}, if you want --- should look like?} I guess up to now nobody 
knows a detailed answer to this. However, as any theoretical physicist 
should be able to answer the question about the fundamental forces of 
nature, he or she should also have at least an opinion about the question 
asked above. 

To break any discussions, here is my coarse answer: {\it A unifying theory} 
should have the {\it potential} to explain the wealth of theoretical models 
and experimental facts within one framework using as few as possible free 
and unconstrained parameters. However, {\it the unifying theory}, i.e.
{\it the TOE}, should in addition show that there is only a {\it unique} 
way in doing this, implying that there are no free parameters at all. 

In the last decades one approach has been proven to be most successful in 
describing our physical world with its fundamental particles and interactions,
the approach summarized within the magic words {\it Quantum Field Theory} (for
a modern introduction see \cite{Wei95a}). Its success is based upon (at least) 
two major concepts: 
\begin{enumerate}
\item {\it The concept of symmetry.} \\
      On the classical level nature is described by classical mechanics,
      quantum mechanics (one should not worry about the word ``quantum'' 
      in this context, it is merely a convention to view the first quantized 
      quantum mechanics as a ``classical'' theory) as well as classical field 
      theory (like the Maxwell theory or Einstein's gravity). The mathematical 
      equations of all these theories are invariant under space--time 
      symmetries, expressing the fact that it should be unimportant when, 
      where or in which position one performs a physical experiment. 
      Moreover, the building blocks of field theories are classical fields 
      which can, in addition, transform under certain global symmetry 
      transformations mixing their internal degrees of freedom, but leaving 
      the underlying equations also unchanged. When this global symmetries
      are made local, the theory in question takes the form of a gauge
      theory. Here new fields (gauge fields) appear acting as transmitters
      of the interaction between the fields which now represent particles. 
      Over the path of second quantization one finally arrives at a 
      quantum field theory. \\
      This way opens up the possibility to describe the electromagnetic, 
      weak and strong interactions within a unique mathematical framework. 
      Without doubts, the corresponding quantum field theories (basing upon 
      the gauge symmetry groups $U(1)$, $SU(2)$ and $SU(3)$, respectively)
      are very successful in describing experimental results (at least in 
      \enlargethispage{0.3cm} the electromagnetic sector) and deepened our 
      understanding of nature.
\item {\it The concept of convergence.} \\
      However, the application of certain quantum field theories to physical 
      problems is mainly based upon one mathematical tool, namely the power 
      expansion with respect to a small parameter. The role of this parameter 
      is taken by the coupling constant whose meaning can be viewed in two 
      ways. First, it describes the strength of the interaction in question. 
      In the formal perturbation expansion the coefficients are given by sums 
      over corresponding Feynman graphs. To be a sensible theory which gives 
      an approximation of any desired accuracy --- only this way a comparison 
      with experiments will be possible and meaningful --- this power series 
      must converge. However, this only happens if the coupling is sufficient 
      weak. Second, the coupling constant can be viewed as describing the 
      strength quantum fluctuations of the fields modifying the free 
      (classical) theory. Only if this fluctuations are sufficient small
      one can trust the results of calculations. 
\end{enumerate}

But despite the great success in describing physical processes within the 
framework of up-to-date quantum field theory (compare the theoretical and
experimental results related to the electromagnetic force), there appear 
more and more rocks on the way to a deeper understanding of such phenomena, 
rocks which become bigger and bigger and cannot be simply thrown aside. 
Especially the weak--coupling requirement sets substantial limits to the 
calculation and treatment of physical effects. One of the most cited examples 
is the quantum field theory of strong interactions, quantum chromodynamics, 
whose treatment in the framework of perturbation theory has born only few 
fruits due to its large coupling. On the contrary, fundamental phenomena 
like confinement are addressed to non--perturbative phenomena meaning that 
their treatment within ordinary perturbation theory will not be possible. 
M.J.Duff summarized this with the words \cite{Duf98a} 
{\it `` `God does not do perturbation theory'; it is merely a technique 
dreamed up by poor physicists because it is the best they can do.''} 
There is nothing more to say! 

However, in addition to this principal problem --- which does not automatically 
imply that current day quantum field theory is wrong; it only implies that up
to now we do have a better way to describe nature --- there is a more substantial 
problem, namely the occurrence of divergencies in certain diagrams of the 
perturbation series. The solution came early in the 1930s and 1940s when 
it was shown that by reshuffling and resumming different Feynman graphs 
in quantum electrodynamics one can cancel this infinities. This procedure, 
which is now known as renormalization, also works for other theories, and 
renormalizability now became one of the main conditions a serious quantum 
field theory has to fulfil. But more to this a little bit later.
 
\begin{figure} 
\centerline{\epsfig{file=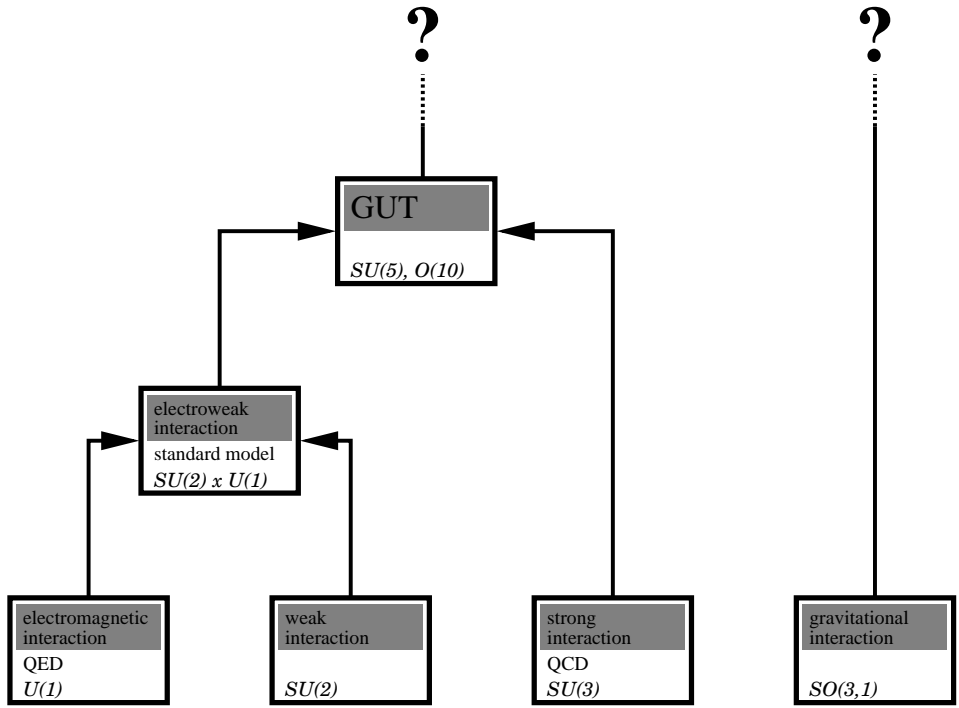,width=5in}}
\caption{\footnotesize
The standard way of unifying the fundamental forces of nature. However, 
the gravitational interaction does not fit into the scheme due to problems 
arising when gravity is treated as a quantum theory.
\hfill
\label{F-Unification}}
\end{figure}

Let us return to the quest for an unifying theory! Despite the problems mainly 
concerned with our limited ability to go beyond the perturbative level, the first 
of the above mentioned basic concepts of QFT, namely the concept of symmetry, 
opened up the way to unify three of the four known fundamental forces of nature.
The receipt is very simple: Find a larger gauge group which contains the gauge 
groups of the known quantum field theories as subgroups, and construct a new
quantum field theory using the common scheme. This receipt has been proven 
to be very successful, at least at a formal level. Besides the fact that this 
way, i.e. by enlarging the gauge group, some of the divergencies occurring 
in the ``smaller'' theories can be eliminated, the standard model (see 
Fig.\ref{F-Unification}) of S.L.Glashow, S.Weinberg and A.Salam (for a 
review see \cite{Lan81a}) became a nearly full accepted theory, although 
some of the main ingredients --- for instance Higgs bosons --- are still lacking 
in the laboratories of the experimentalists. Other standard model puzzles, like 
the fermion masses and charge quantization, are solved to more or less 
satisfaction within the framework of even ``larger'' theories, like the Grand 
Unified Theories (see for instance \cite{BurEllGaiNan78a}), although even here 
new problems arise.

Nevertheless, without going further into details, one is faced with at least two 
subtleties on this way of unification. First of all it is strictly based upon 
physical experience. Of course the agreement with everyday physics is the last 
(or first?) ``external consistency'' check for a theory. But in the above 
mentioned cases of electroweak and grand unified models the outputs of physical 
experiments --- namely the fact that {\it there are} electromagnetic, weak and strong 
forces in the nature (I will exclude gravity for the moment) --- are the 
fundamental building blocks. What happens if some experimentalists measure
effects of a very weak fifth force whose quantum field theoretical model does not 
fit into the unification scheme used up to this point? There are no ``internal'' 
or theoretical arguments showing that the fundamental forces known today are the 
only possible or realized ones. In fact, I believe that in the framework of the 
common unification scheme such arguments will never be available!

This brings me straight to the second subtlety, which concerns the richness 
and arbitrariness of possible unifying models: The most popular choices for 
the ``grand unified'' gauge group --- $SU(5)$ or $O(10)$ --- are the simplest 
ones (hence the name ``minimal models''). Much more complex structures are
conceivable. Of course, in addition to the ``external consistency'' checks 
mentioned above, each of the conceivable models has to obey also ``internal
consistency'' checks (meaning that their mathematical structure and physical 
interpretation should not lead to contradictions). But this is only of little
help for reducing the infinite number of possible grand unified models.

With this in mind, one has to recall the question asked at the beginning in a 
slightly modified form: {\it Does the way of unification described so far 
open up a way to a grand unified theory?} Yes, without doubts! But with 
emphasise lying on the words {\it a grand unifying theory}, i.e. a theory 
which will be only one in the huge space of all possible and allowed unifying 
models. Moreover, using the receipt sketched above, one does not leave the common 
framework of QFT. Thus, besides the just mentioned subtleties of more 
philosophical nature, we are again faced with all other problems of more 
mathematical nature appearing in the ordinary quantum field theoretical 
approach. This brings us to the question, whether the description of physical
phenomena within the framework of formal QFT and its concrete utilization in 
perturbation theory provides the correct way to a deeper and more fundamental 
understanding of nature going beyond the current knowledge?

To make this way of thinking more clear --- and to demonstrate that 
this can be viewed as the birthplace of string theory, at least in a 
``philosophical'' sense --- let us take some closer look at quantum field 
theories and their realization. As mentioned above, within the framework of
perturbation theory concrete results are deviated using a formal power 
expansion with respect to a (hopefully small) coupling constant. The 
coefficients of the resulting series are sums over appropriate Feynman 
diagrams. Those diagrams --- although very useful and illustrative --- bear 
the first danger, namely, they are just graphs and not manifolds. With 
other words, at an interaction point the local topology is not $\R^n$. 
This fact has at least two significant effects. First, there is no restriction 
to introduce at that interaction points arbitrarily high spins because there 
is no correlation between the internal lines (indicating the propagation 
of quantum particles) and the vertices (indicating points of interaction 
between the quantum particles). That is, point particle theories are accompanied 
by the problem of an infinite degree of arbitrariness concerning the structure 
of possible interactions. Of course, demanding renormalizability sets strict 
constraints to the type of interactions, but, nevertheless, the possibility 
to write down an infinite number of renormalizable quantum field theories 
for point particles remains. The second point is, that Feynman diagrams with 
at least two interaction points are suffered from ultraviolet divergencies 
which occur if we ``pinch'' the diagram by shrinking an internal line to 
zero and, thus, deforming the local topology of the graph. The occurring 
infinities can in some cases be removed by renormalization. 

\begin{figure} 
\centerline{\epsfig{file=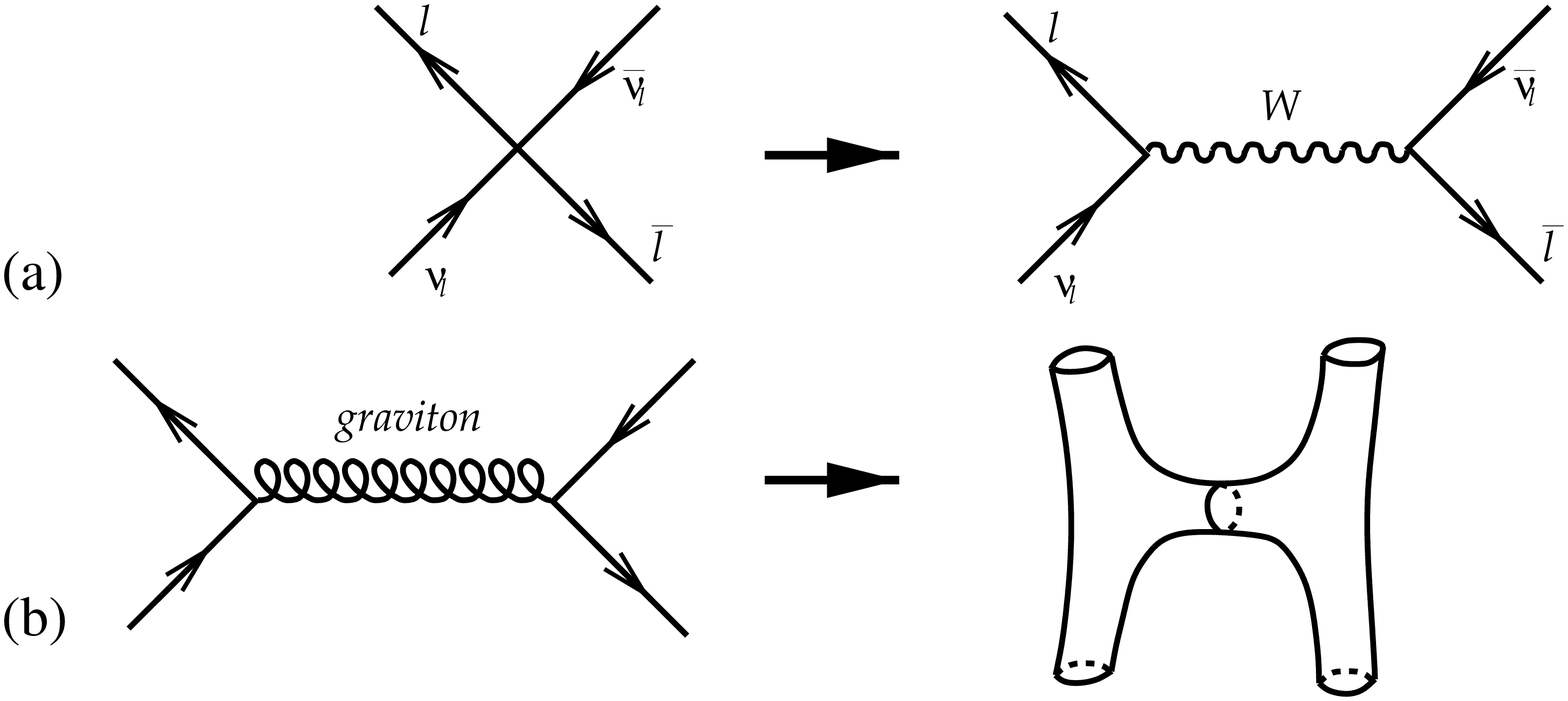,width=4in}}
\caption{\footnotesize 
{\it New physics} appearing when one goes to some higher energy scale.
Figure (a) shows how divergencies occurring in the leptonic weak interaction 
of four--Fermi theory can be removed by resolving the contact interaction at 
short distances (i.e. high energies) into the exchange of a corresponding $W$ 
boson. Similar, the potential infinities appearing at high energies due to 
the exchange of a graviton between two point particles shown in figure (b) 
can be resolved by replacing the point particles with one--dimensional 
objects (strings). This provides a natural cut--off scale and, thus, removes
automatically all ultraviolet divergencies.
\label{F-Smooth}}
\end{figure}

However, things become even worse if we try to unify gravity (which up to 
this point was exclude from the discussion) and quantum mechanics, i.e. 
if we try to build up a consistent QFT of gravity (for a recent review of 
present--day approaches see \cite{Rov98a}). Here a renormalization 
procedure does not provide a way out of the appearing difficulties, 
especially the so--called ``short distance problem''. The reason 
for this bad behavior is related to the fact that gravity couples to energy 
rather than charge, and that the gravitational coupling (i.e. Newton's 
constant) in natural units is given by $g_N = l_P^2 = m_P^{-2}$ (where 
$l_P \sim 1.6 \times 10^{-33} cm$ denotes the Planck length and 
$m_p \sim 10^{-8} kg \sim 10^{19} GeV$ the Planck mass), thus 
having the dimension $[length]^2$ or $[mass]^{-2}$. For instance, 
the one--graviton correction to the original gravitational scattering 
amplitude of two point particles (see first picture in 
Fig.\ref{F-Smooth}(b)) is proportional to Newton's constant and the 
square of the typical energy $E$ involved in the process, $E^2/m_P^2$. This 
dimensionless ratio indicates that the strength of this correction is small 
at long distances, i.e. low energies, but becomes large at higher energies 
$E>m_P$. This way the perturbative power expansion with respect to 
quantum corrections due to graviton exchange becomes useless at short 
distances, meaning that perturbation theory breaks down. 

As already mentioned above, a renormalization procedure, which eliminate 
unwanted infinities by an infinite redefinition of the parameters of the 
theory, does work very well for the known quantum field theoretic models 
of strong, weak and electromagnetic interactions, but cannot be applied 
to quantum gravity. Here we have a power expansion in a dimensional 
parameter $g_N$ and, therefore, we are no longer able to reshuffle and 
resumme graphs at different order in the power series to cancel the 
occurring infinities. Thus, the ordinary (naive) renormalization theory 
does not work, making a quantum field theory of gravity a non--renormalizable 
theory.

Of course there are several other ways which may lead out of this problems. 
One path is given by putting calculations on a lattice. I do not want to 
argue this way any further. Just let me stress that, although it can be very 
useful, in my opinion the lattice is nothing but a tool to remove substantial 
difficulties occurring in a concrete calculation; principal problems of 
the underlying continuum theory coming from a limited understanding or 
the limits of the used model itself, cannot be solved this way --- they 
can only be thrown to another place. Moreover, in the case of gravity we 
know that Lorentz invariance holds to a very good approximation in the 
low energy theory. Thus, if we go on the lattice by making the interaction 
in the spatial directions discreet, we have to do this at the same time as 
well in the time direction. This will result in the loss of causality and 
unitarity, two of the basic demands which a consistent theory has to fulfil.
 
A second path is given by supersymmetry whose study began in the 1970s as 
the only possible extension of the known space--time symmetries of particle
physics (for an short introduction with emphasis on some modern aspects see
\cite{Lyk96a}) and which may lead to a unifying theory beyond the 
standard model. As remarked above, an enhancement of symmetries by introducing 
larger gauge groups in a gauge field theory can cancel a large class of 
divergencies due to Ward--Takahashi identities. Despite the fact that the 
supersymmetric partners of the known particles are still waiting for their
observational discovery, this way it is possible to construct quantum theories 
of gravity that are finite to every order in the coupling constant. Such models 
are called supergravity theories. However, popular models in four dimensions 
(such as the $O(8)$ or $Osp(N/4)$ model) are either to small to accommodate 
the minimal $SU(3)\times SU(2) \times U(1)$ grand unified model or too small 
to eliminate all occurring divergencies. 

However, a third and maybe the most ``natural'' path --- at least from the 
viewpoint of a string theorist --- is to interpret the point particle quantum 
field theories as valid only up to some energy scale beyond which one is 
faced with {\it new physics} \cite{Pol96a}. One way to look at this 
{\it new physics} is that it should have the effect of smearing out the 
interaction in space--time. Thus, by providing a natural ultraviolet 
cut--off scale, this will soften out the high energy behavior of the 
theory in question. However, the number of possible ways on which 
this could be achieved is very limited because the combined constraints 
of Lorentz invariance and causality set profound restrictions.

Let me make this viewpoint more clear by considering the following simple 
example. In the four--Fermi theory the leptonic weak interaction (see 
Fig.\ref{F-Smooth}(a)) leads to divergencies which can be removed in the 
context of Weinberg--Salam theory by resolving the contact interaction at 
short distances --- and thus high energies --- into the exchange of a
corresponding $W$ gauge boson of weak interaction. Due to internal 
consistency reasons, this is the only way to solve the short--distance 
problem of weak interaction. In a similar way (see Fig.\ref{F-Smooth}(b)),
the exchange of gravitons between two elementary particles, which also 
leads to potential infinities at high energies due to the short--distance 
problem when described within the framework of usual QFT (see above), can 
be resolved by ``smearing out'' the point particle, for instance by 
replacing the point particles with one--dimensional objects --- 
{\it strings}. The Feynman graphs get replaced by a smooth two--dimensional 
genuine manifold (called world--sheet) which one cannot shrink to zero 
without changing the whole topology of the corresponding diagram. Thus, 
one obtains a ``natural'' cut--off preventing ultraviolet divergencies;
one cannot ``pinch'' the world--sheet of a string to obtain an ultraviolet 
divergence due to topological reasons.

This way we are faced for the first time --- although in a very naive way --- 
with the concept of strings. But the replacement of point particles by 
one--dimensional objects has even more consequences as the simple resolution 
of ultraviolet divergencies. Some of these will be the topic of the next 
section. Moreover, there we will recall the original question whether 
this new viewpoint provides an acceptable and maybe deeper answer to the 
unifying problem of physics bringing us beyond the usual approaches.

\section{String Basics or {\it What are Strings?}}
\label{StringBasics}

Curiously, the original approach to string theory was addressed neither to 
the solution of the short--distance problem of quantum gravity or the 
corresponding ultraviolet divergencies, nor the search for a unifying 
theory. It was discovered more unexpected in the late 1960s when Y.Nambu, 
H.B.Nielsen and L.Susskind made the remarkable observation that the 
Veneziano model, constructed to explain the abundance of hadronic 
resonances in experiments, describes the scattering of 
one--dimensional objects, strings. Up to the mid 1970s this new theory 
was studied as a possible theory of strong interaction. However, a closer 
look revealed a lot of unwanted features because Lorentz invariance sets
high constraints to make it a consistent theory. So it contained besides 
tachyons with negative mass--squared a mysterious massless spin--two particle. 
Despite this, the most striking fact was that the model could seemingly 
only be consistent in specific dimensions, called ``critical dimensions'', 
which were found to be 26 for the Veneziano model and 10 for the later 
invented Ramond--Neveu--Schwarz model incorporating in addition to bosons 
also fermions.

Thus, the idea of string--like particles was dropped out of the physicists 
minds and forgotten for about one decade. But after the first euphoric 
successes of ordinary QFT it became more and more clear, that even here
one is faced with many new and difficult problems. One of these was the 
observation that it seemed to be impossible to treat the last of the four 
fundamental forces, gravity, within a quantum theoretical framework. 
Each attempt to construct a quantum gravity led to a dead--end. But
some physicists remembered a formerly unwanted feature of the forgotten 
string theory, namely that it provides from the very beginning a massless 
spin--two particle which could be viewed as the graviton --- the transmitter 
of gravitational force. Every consistent string theory predicts gravity, 
i.e. every string theory includes ``automatically'' a quantum theory of 
general relativity! Moreover, further calculations showed that the quantum 
corrections to amplitudes are ultraviolet finite to all orders. So 1984, 
among others, M.B.Green and J.Schwarz continued to study string theory, 
this time as a leading candidate for a unification of QFT and gravity. 

Then, in the second half of 1984, happens what now is called the {\it first 
superstring revolution}: M.B.Green and J.Schwarz showed that a special type
of string theories, called type $I$ string theory, is free from anomalies 
if the ten--dimensional gauge group is uniquely $SO(32)$. D.Gross, J.Harvey, 
E.Martinec and R.Rohm discovered a new consistent kind of heterotic (hybrid) 
string theory based on just two groups: $E_8 \times E_8$ and $SO(32)$. 
Last but not least, P.Candelas, G.Horowitz, A.Strominger and E.Witten 
showed that these heterotic theories admit a Kaluza--Klein compactification 
from ten to four dimensions leading to the known grand unifying model of 
strong, weak and electromagnetic forces. A great success which showed 
that string theory is much more than just an exotic idea of some theorists!

\begin{figure} 
\centerline{\epsfig{file=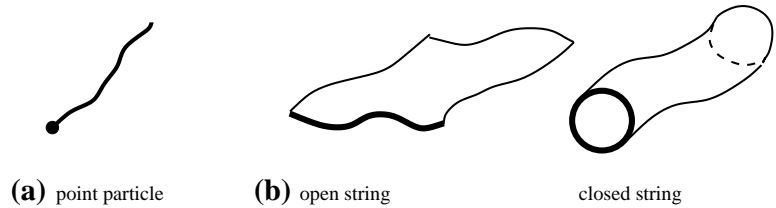,width=4in}}
\caption{\footnotesize 
The point particle of ordinary quantum field theory (a) sweeping 
out an one--dimensional world--line gets replaced in string theory
by one--dimensional objects of Planck--length (b) sweeping out 
a two--dimensional world--sheet in a higher dimensional 
target--space.
\label{F-Strings}}
\end{figure}

So, what is string theory? At this time it was ``nothing but'' the quantum 
theory of one--dimensional extended objects, with a length of about 
$10^{-33} cm$ --- thus far away to be resolved by present day experiments --- 
and only one free parameter, namely the energy per unit length (string 
tension). These strings move in a higher--dimensional target space and 
sweep out two dimensional surfaces called world--sheets, see 
Fig.\ref{F-Strings}. One distinguishes between open and closed strings, 
leading in the target space to a long strip of finite width or a long tube, 
respectively. This way string theory can be considered as a conformal field 
theory (a two--dimensional quantum field theory) where the spatial
direction labeling the coordinate on the string is finite. With that in mind, 
conformal field theories, formerly studied merely as simple toy models, now 
become the mathematical framework for a realistic theory of unification.

Particles in the target space --- and, thus, the particles moving in 
space--time --- are identified with the various eigenmodes of the string. 
That is, they are nothing but excitations of one fundamental string, just 
like the vibration modes of a violin string produce various sounds. To be 
more concrete, the massless modes, which correspond to the lowest excitation, 
lead to the particles contained in the standard model (i.e. gauge bosons, 
leptons, quarks), whereas the higher excitations produce an infinite tower 
of heavy particles with masses of order of the Planck mass (i.e. $10^{19} GeV$) 
which, for that reason, will be unobservable in current--day experiments. 

However, conformal invariance, modular invariance and cancellation of 
anomalies in addition to the already mentioned Lorentz invariance puts 
substantial restrictions on the number of space--time dimensions of the 
target space as well as the possible space--time particle spectrum. As in 
the case of the old Veneziano model, a consistent formulation of string 
theory including bosonic and fermionic excitations (superstrings) is
only possible in a ten--dimensional target space. But despite this
drawback it turns out, that in this ``critical dimension'' there 
are only {\it five} consistent superstring theories --- type $IIA$, type
$IIB$, type $I$, type $SO(32)$ heterotic and type $E_8 \times E_8$ heterotic
string theory --- differing in properties like the type of the involved 
strings (open or closed, oriented or non--oriented), the field content of 
the world--sheet theory, the number of supersymmetry generators or the 
ten--dimensional gauge group. The following table and list summarize some 
of that properties (for much more information there exists a plenty of good 
literature about strings and related fields; here I want to mention only the 
standard book about strings --- {\it Superstring Theory} by M.B.Green, 
J.Schwarz and E.Witten \cite{GreSchwWit87a} --- and the latest book --- 
{\it String Theory} by J.Polchinski \cite{Pol98a} --- which contains some 
of the most recent developments). \\
\ \\
{\small
\begin{tabular}{|p{2cm}|p{1.7cm}|p{1.7cm}|p{2.2cm}|p{2.2cm}|p{2.2cm}|}
\hline 
		      & $IIA$		       & $IIB$			& $I$					& $SO(32)$ heterotic string & $E_8 \times E_8$ heterotic string \\
\hline
stringtype	      & closed                 & closed         	& non--oriented                 	& closed 	            & closed \\
                      & oriented               & oriented               & open and                              & oriented                  & oriented \\
		      &                        &                        & closed                                &                           & \\
\hline
ten--dim. 	      & $N=2$                  & $N=2$   		& $N=1$ 				& $N=1$ 		    & $N=1$ \\
SUSY                  & non--chiral            & chiral                 &                                       &                           & \\
\hline
ten--dim. gauge group & none		       & none			& $SO(32)$				& $SO(32)$ ($Spin(32)/\Z_2$)& $E_8 \times E_8$ \\
\hline
\end{tabular}

\begin{enumerate}
\item {\bf type $IIA$ string theory:} \\
      {\it field content of the world--sheet theory:} world--sheet theory 
        is a free field theory containing 8 scalar fields 
        (representing the 8 transverse coordinates of a string moving in 
	nine spatial directions) and 8 Majorana fermions (regarded as 16 
	Majorana--Weyl fermions, 8 of them having left--handed chirality and 
	the other 8 having right--handed chirality) \\
      {\it massless bosonic spectrum:}
	\begin{itemize}
	\item $NS-NS$--sector: metric, antisymmetric tensor, dilaton 
	\item $R-R$--sector: vector potential, rank three antisymmetric tensor
	\end{itemize}
      {\it remarks:}
	\begin{itemize}
	\item the 8 scalar fields satisfy periodic boundary conditions; 
	      the fermions are chosen to obey either periodic (Ramond, $R$) 
	      or anti--periodic (Neveu--Schwarz, $NS$) boundary conditions	
	\item type $IIA$ string theory seems to be non--realistic
	      in the sense that no realistic QFT in lower dimensions can be 
	      deduced from it
        \end{itemize}

\item {\bf type $IIB$ string theory:} \\
      {\it field content of the world--sheet theory:} world--sheet theory 
	is a free field theory containing 8 scalar fields and 8 Majorana 
	fermions (see type $IIA$ string theory) \\
      {\it massless bosonic spectrum:}
	\begin{itemize}
	\item $NS-NS$--sector: metric, antisymmetric tensor, dilaton 
	\item $R-R$--sector: rank four antisymmetric tensor gauge field 
	      satisfying the constraint that its field strength is self--dual,
	      rank two antisymmetric tensor field, scalar
        \end{itemize}
      {\it remarks:} 
	\begin{itemize}
	\item as in type $IIA$ string theory the 8 scalar fields satisfy 
	      periodic boundary conditions, whereas the fermions can have
	      Ramond or Neveu--Schwarz boundary conditions	
	\item seems to be non--realistic due to the same reasons as 
	      type $IIA$ string theory
	\end{itemize}
      
\item {\bf heterotic $E_8 \times E_8$ string theory:} \\
      {\it field content of the world--sheet theory:} world--sheet theory 
	consists of 8 scalar fields, 8 right--moving Majorana--Weyl fermions 
	and 32 left--moving Majorana--Weyl fermions; 
	NS and R boundary conditions for the right--moving fermions \\
      {\it massless bosonic spectrum:}
	metric, antisymmetric tensor field, dilaton, set of 496 gauge fields 
	in the adjoint of $E_8 \times E_8$ \\
      {\it remarks:}
	leads to four--dimensional theories which resemble quasi--realistic 
	grand unified theories with chiral representations for quarks and 
	leptons
      
\item {\bf heterotic $SO(32)$ ($Spin(32)/\Z_2$) string theory:} \\
      {\it field content of the world--sheet theory:} world--sheet theory 
	consists of 8 scalar fields, 8 right--moving Majorana--Weyl fermions 
	and 32 left--moving Majorana--Weyl fermions; NS and R boundary 
	conditions for the right--moving fermions \\
      {\it massless bosonic spectrum:}
	metric, antisymmetric tensor field, dilaton, set of 496 gauge fields 
	in the adjoint of $SO(32)$ \\
      {\it remarks:}
	leads also to four--dimensional quasi--realistic grand unified 
	theories with chiral representations for quarks and leptons
      
\item {\bf type $I$ string theory:} \\
      {\it field content of the world--sheet theory:} world--sheet theory 
	is a free field theory containing 8 scalar fields and 8 Majorana 
	fermions \\
      {\it massless bosonic spectrum:}
	\begin{itemize}
	\item $NS-NS$--sector: metric, dilaton 
	\item $R-R$--sector: rank 2 antisymmetric tensor field 
	\item open string sector: set of 496 gauge fields in the adjoint 
	      of $SO(32)$
        \end{itemize}
      {\it remarks:}
	\begin{itemize}
	\item $SO(32)$ Chan--Paton factors coupling to the ends of the 
	      open string
	\item looks realistic as a unified theory in the sense that it 
	      incorporates internal symmetry groups containing the 
	      $SU(3) \times SU(2) \times U(1)$ of the common grand unified
	      model
	\end{itemize}
\end{enumerate}
} 

This was a indeed great surprise! The infinity of consistent point particle 
theories in four dimensions on one side faces only five consistent string 
theories in ten dimensions on the other side. What a big step towards a 
unification! However, at least two points were still open in the mid 1980s
and determined significantly the further way of string theory. First, there
was the question how to step down from ten dimensions to the real four 
dimensional world. The second open problem concerned the application of
string theory, i.e. the question how to get experimental verifiable 
results, or at least results which can be compared with ordinary quantum
field theory. 

\begin{figure} 
\centerline{\epsfig{file=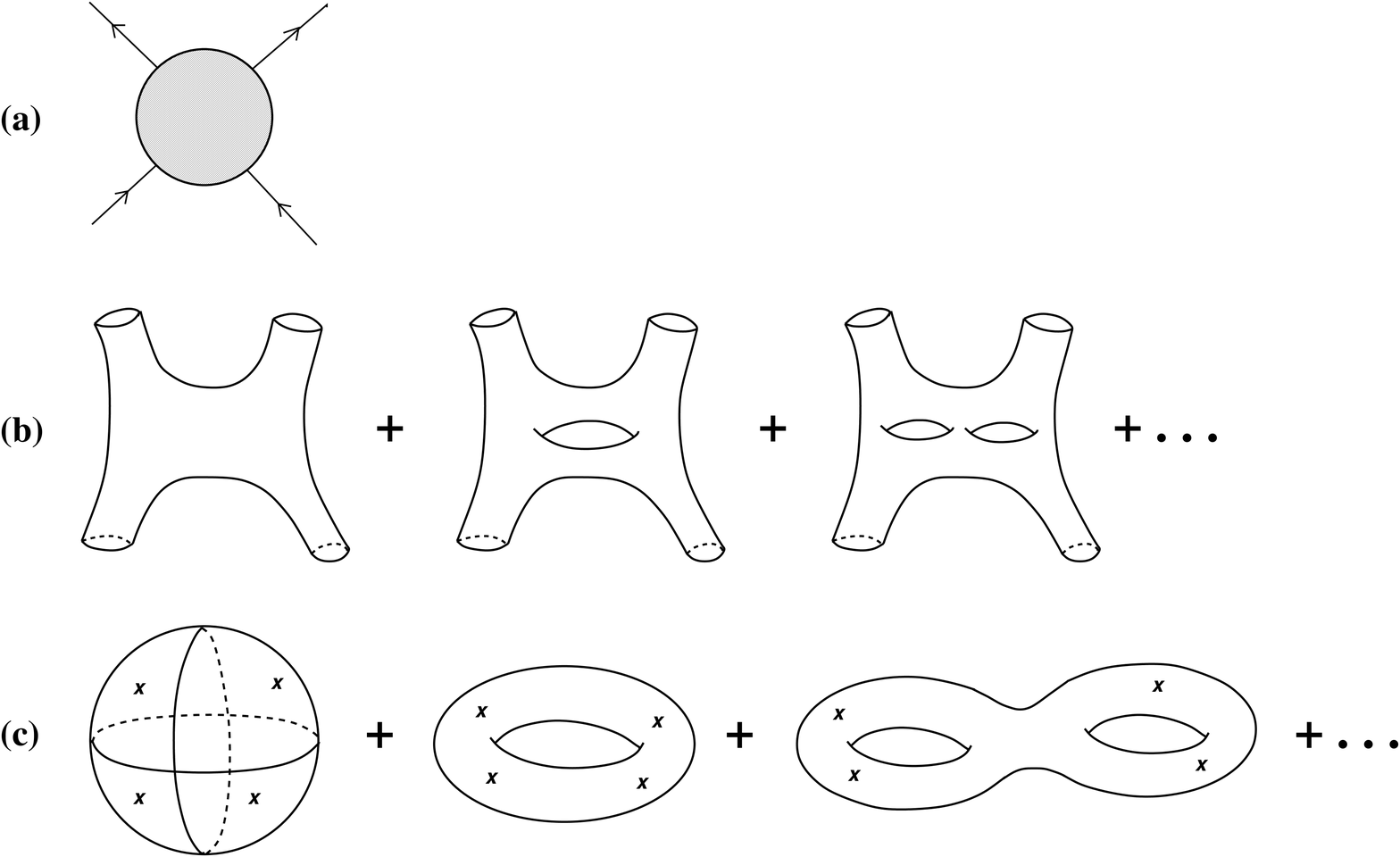,width=4in}}
\caption{\footnotesize 
From point--particle scattering to string-string scattering. Figure 
(a) shows the scattering of two particles in ordinary QFT. The shaded 
region denotes the infinite number of Feynman graphs one has to sum up.
One problem arising within this approach is that the number of Feynman
graphs increases very rapidly with increasing order, making the calculation
of higher order corrections more and more difficult. The picture changes in 
string theory, figure (b). Here there is only one diagram per order, where 
the latter is given by the topological genus of the corresponding Riemannian
surface. Finally, using the conformal properties of string theory, the 
calculation of the two particle scattering amplitude can be mapped to the 
sum over compact Riemannian manifolds labelled by their topological genus,
where the incoming and outgoing particles are represented by appropriate 
vertex operators inserted at the surfaces, see figure (c).
\label{F-Scattering}}
\end{figure}

The latter is handled within the framework of string perturbation theory, 
where the far--reaching and well understood topics of Riemannian geometry
come into play. This perturbative approach to string theory is mainly based
upon the thoughts of ordinary path integral methods --- maybe because this
is just a first try or due to the lack of any better way. To be more concrete,
the string scattering amplitudes are defined as path integrals over the 
two--dimensional quantum field theory on the world--sheet, with insertions 
of suitable vertex operators representing the particles being scattered 
(see Fig.\ref{F-Scattering}). This corresponds to the calculation 
of correlation functions of vertex operators in two--dimensional conformal
field theories. That recipe takes automatically into account the infinite 
number of massless and massive modes which can be exchanged in the 
scattering process. Thus, whereas in the calculation of a corresponding 
scattering amplitude within the framework of ordinary point particle QFT 
all different Feynman graphs has to be taken into account (whose number 
become very large in higher orders), in string perturbation theory there 
is only one diagram per order. Here the order is given simply by the 
genus of the corresponding Riemannian surface. Potential problems 
arise because the topological sum is augmented by integrals over 
conformally equivalent shapes that the Riemannian surface at a given order 
can have. Indeed, the ``correct'' integration variables are only those 
describing conformally inequivalent surfaces which are mathematically 
hard to describe. This fact makes calculations beyond the one--loop level 
very difficult. Nevertheless, the genuine absence of ultraviolet 
divergencies to all orders of the perturbation series was such a 
strong motivation to overlook this difficulties.

The second and much more potential drawback arose from the question, how to 
descend from the giddy highs of ten dimensions to the real four--dimensional
world or, with other words, how to construct low energy ``effective'' theories.
The magic word which comes into play is {\it compactification}: Dismantle
the ten--dimensional target space into a four--dimensional ``large'' piece 
building up our real world, and some compact six--dimensional manifold ``small''
enough to vanish at low energies from our view. The good news are, that 
indeed such compactification schemes exists which lead at low energies to 
theories exactly of the type wanted, namely the grand unifying model of strong, 
weak and electromagnetic forces. However, the bad news are, that there are 
possibly an infinite number of ways to compactify the ten--dimensional string 
theories, leading to a plenty of different lower--dimensional theories and,
thus, predictions for a ``real world''. This substantial difficulties are now 
summarized as the {\it vacuum--degeneracy problem}: the parameters or moduli
which determine the shape and properties of the compact manifold correspond 
``physically'' to vacuum expectation values of scalar fields in the string 
theory. During the compactification procedure they enter the obtained
effective field theories in lower dimensions as free parameters which are 
not constrained by any obvious fundamental principles. In contrast, they build 
up what is called moduli space or vacuum. This way, their values are only 
determined by choosing --- merely ``by hand'' --- a special vacuum or point 
in this moduli space. On the other hand, these parameters are related to the 
physical observables of the effective field theories, like mass parameters, 
coupling strengths of the interactions and so on. With other words, most of 
the properties of the low energy theory are not determined by the microscopic 
theory itself, i.e. the terms entering in the Lagrangian, but by the choice 
of the vacuum structure --- and, unfortunately, there are infinitely many ways 
in doing this. The main problem is that there is no dynamical mechanism that 
would single out a particular vacuum.

But recall the question asked at the beginning, namely the question about the 
nature of a unifying theory, and let us summarize what we have learned so far. 
There are only five consistent string theories in ten dimensions with only 
one free parameter, the string tension. Moreover, internal consistency 
dramatically reduces the number of possible low energy spectra and independent 
couplings --- at least compared to ordinary quantum field theories. Despite 
many difficulties it was shown that three of this five string theories contain 
the standard model of strong, weak and electromagnetic forces, describing our 
real world, as a subtheory. Thus, string theory has at least the {\it potential} 
to explain the things that happen in the laboratories of the experimentators, 
making it without doubts {\it a unifying theory}. However, besides others, the 
vacuum--degeneracy problem is one of the huge rocks blocking the way to make 
string theory much more like this, namely {\it the unifying theory} going 
beyond the current understanding of nature. Even this fact has led --- for the 
second time --- to the retreat of string theory from an active field of research 
at the end of the 1980s. Moreover, it splits the physical community into two 
nearly disjunct groups --- one group whose members think that those huge 
rocks could never be thrown out of the way, and a second group of unshakeable
enthusiasts who believed and still believe in the potential strength of 
string theory.

\section{The Future of Strings or {\it Quo Vadis?}}
\label{FutureStrings}

The faith of those string enthusiasts became more and more substantial by one 
fundamental consideration: There are five consistent superstring theories in 
ten dimensions. Indeed, this number is much lower than the number of consistent 
point particle quantum field theories in four dimensions, but still too large 
for the one and only unifying theory. But before choosing one of the string 
theories to be the ``most fundamental'' one and asking the question about the 
meaning of the others, one should spend some time to think about the question 
in which sense these five string theories are different. A possible answer 
comes immediately! The formulation of the five superstring theories is based 
upon the tools of up--to--date QFT, which is only successfully accessible in 
the framework of perturbation theory. Thus the question arises, if the five 
different but consistent string theories are merely a result of their
perturbative formulation, i.e. the standard way calculations are carried out 
in QFT, rather than a result of a ``deeper'' lying and not yet understood 
principle which says that there has to be five fundamental theories? 

Even here the answer became quickly clear at the beginning of the 1990s: 
{\it Duality}. This concept was not new to the physical community. 
Already in 1931 P.Dirac pointed out the invariance of Maxwell's equations 
under the exchange of the electric and magnetic field strength after the 
introduction of magnetic sources. Over 40 years later, in 1977, C.Montonen 
and D.Olive showed that this electric/magnetic (or strong/weak) duality is 
indeed an exact symmetry of the whole QFT. However, although very simple, 
the idea of duality existed merely as a more hidden playground for 
mathematical physicists; up to the late 1980s when first indications 
arose within the context of string theory that duality is much more 
than just a really nice looking but more or less useless symmetry 
of the considered field theories. The more light were thrown into the dark, 
the more it became clear that this special concept of symmetry may be one of 
the leading principles of nature (especially from a unifying viewpoint),
just like its ``big'' or ``little'' --- depending from the viewpoint --- 
brothers, the gauge symmetries of quantum field theories. It induced what 
after the inspiration of J.Schwarz \cite{Schw96a} now is called the 
{\it second superstring revolution}. 

In the last years the investigation of dualities in string theory became one 
of the main research fields. It turned out that there are three fundamental 
types of duality symmetries which are called $S$--duality, $T$--duality and 
$U$--duality (for further information see \cite{Pol96a,Sen98a} and references 
therein):
\begin{itemize}
\item {\it $S$--duality} \\
      This duality describes the quantum equivalence of two theories $A$ and
      $B$ which are perturbatively distinct. Its central idea is, that the 
      strong coupling limit of string theory $A$ is equivalent to the weak 
      coupling limit of string theory $B$ and vice versa, see 
      Fig.\ref{F-SDuality}(a). Thus, perturbative excitations of $A$ are 
      mapped to non--perturbative excitations of the dual theory $B$ and 
      vice versa, making $S$--duality a non--perturbative symmetry. If the
      theories $A$ and $B$ are the same, one has $S$--self--duality, see 
      Fig.\ref{F-SDuality}(b). Examples of this type of duality are the 
      equivalence of type $I$ string theory and $SO(32)$ heterotic string 
      theory, and the self--duality of type $IIA$ string theory.
      
\begin{figure} 
\centerline{\epsfig{file=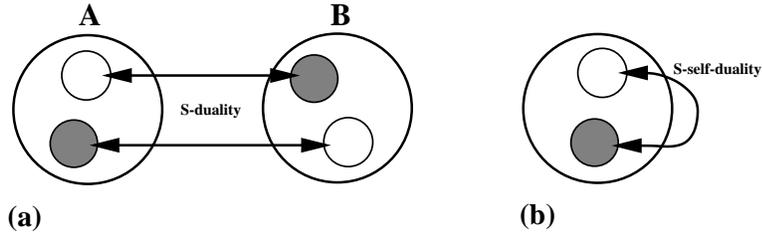,width=4in}}
\caption{\footnotesize 
$S$--duality. Inside the shaded regions the coupling is 
weak and perturbation theory is valid. $S$--duality relates the
weak coupling limit of theory $A$ with the strong coupling limit 
of another perturbatively distinct theory $B$ and vice versa, 
see figure (a). If both theories coincide one speaks of
$S$--self--duality,
see figure (b).
\label{F-SDuality}}
\end{figure}

\item {\it T--duality} \\
      This type of duality relates the weak coupling limit of theory $A$ 
      compactified on a space with large volume to the weak coupling limit of 
      another theory $B$ or $A$ itself (self--duality) compactified on a space 
      with small volume and vice versa. Hence $T$--duality is a perturbatively 
      verifiable symmetry (see Fig.\ref{F-TDuality}). 
      Examples are the duality between type $IIA$ string theory compactified 
      on a sphere $S^1$ of radius $R$ and the type $IIB$ string theory 
      compactified on $S^1$ with radius $R^{-1}$, as well as the self--duality 
      of each of the heterotic string theories when compactified on $S^1$ with 
      radius $R$ and compactified on $S^1$ with the inverse radius $R^{-1}$.

\begin{figure} 
\centerline{\epsfig{file=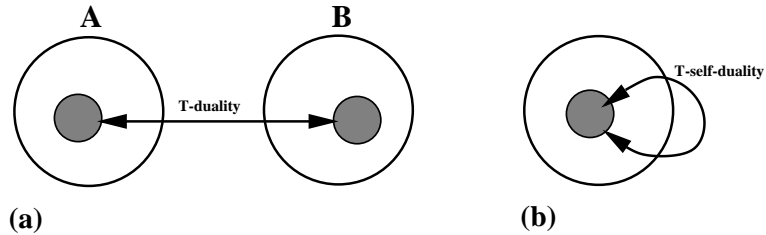,width=4in}}
\caption{\footnotesize 
$T$--duality. Inside the shaded regions the coupling is 
weak and perturbation theory is valid. Figure (a) shows 
how $T$--duality relates the weak coupling regime of 
theory $A$ (compactified on a space with large volume)
to the weak coupling regime of another theory $B$
(compactified of a space with small volume). If both
theories coincide one speaks of $T$--self--duality, 
see figure (b).
\label{F-TDuality}}
\end{figure}

\item {\it $U$--duality} \\
      This third type of duality symmetries combines $S$-- and $T$--duality,
      making theory $A$ compactified on a space with large (small) volume 
      is dual to the strong (weak) coupling limit of another theory $B$. 
      Thus, $U$--duality is also a non--perturbative duality.
\end{itemize}

However, a closer look also reveals one of the current main problems with 
this very powerful tool of duality symmetries, namely the question how to
perform tests of duality conjectures. Because its historical appearance, 
up to date we do not have an independent description of string theories at 
strong couplings. Thus, an exact proof of dualities, especially of $S$-- and 
$U$--duality conjectures, is still lacking. The only way to test it is by 
working out various consequences using certain constraints and symmetries of 
string theory. Up to now there are two main streams which one follows.
The first is the analysis of the low energy effective action, obtained 
from a perturbatively formulated string theory by restricting to the 
lowest lying (massless) excitations. Most of the duality conjectures are 
tested this way by comparing the effective actions of two string theories 
compactified on some manifold. However, although very simple, this method 
provides only a very crude test of duality.

Much more exact are tests involving the spectrum of the corresponding string 
theories. Here the attention lies on a very special part of the spectrum in 
superstring theories, namely the so--called BPS--saturated states  --- 
or, shorter, BPS states --- named after Bogomol'nyi, Prasad and Sommerfeld. 
Such states are invariant under parts of the underlying supersymmetry 
transformation and characterized by two important properties: First, 
the mass of a BPS state is completely determined by its charge as a 
consequence of the supersymmetry algebra. Hence the properties of 
such states are not modified by quantum corrections, independent 
how strong they are. Second, the degeneracy of a BPS multiplet 
is independent of the point chosen in the moduli space. 
With other words, the degeneracy at any value of the string coupling is 
the same as that at weak coupling, making it possible to compare perturbative 
and non--perturbative formulations. This provides a non--trivial test of the 
corresponding non--perturbative duality conjectures. However, the detailed 
mathematical realization of such tests is often very difficult --- but we are 
just at the beginning on the way to an understanding of non--perturbative 
phenomena.

\begin{figure} 
\centerline{\epsfig{file=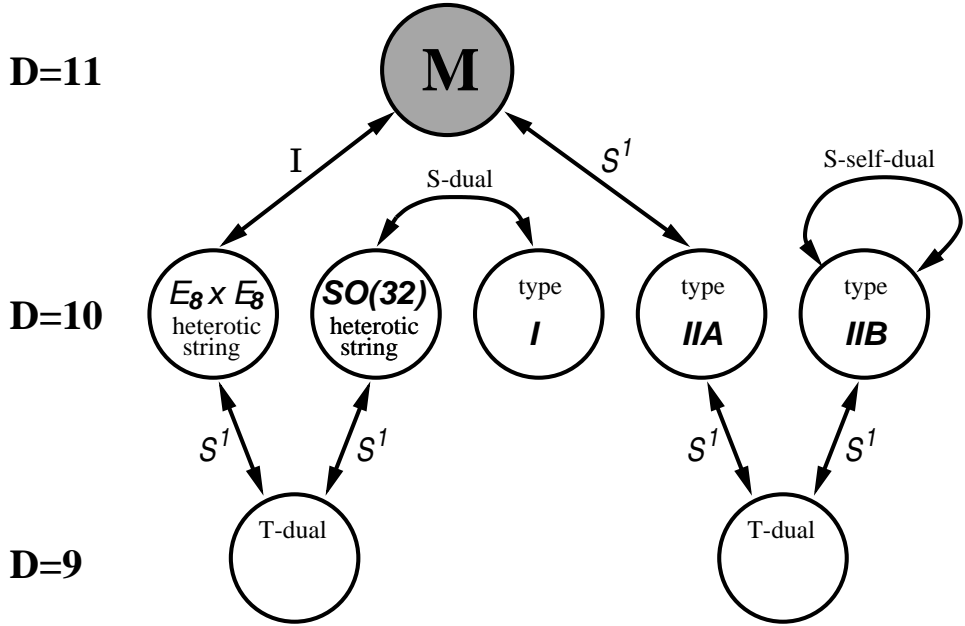,width=5in}}
\caption{\footnotesize 
The net of string dualities in higher dimensions. As it
is shown, the perturbatively different string theories in 
ten dimensions are connected by duality transformations
using certain compactifications. Moreover, in eleven dimensions
a new theory arises, which is called $M$--theory and whose
compactification on the circle $S^1$ or the finite line interval $I$
gives rise to type $IIA$ and $E_8 \times E_8$ heterotic string theory, 
respectively.
\label{F-StringDualities}}
\end{figure}

Nevertheless, despite the lack of rigorous proofs of duality conjectures, the 
``old'' picture of string theory at the end of its first revolution changed 
dramatically when duality pushes some rocks out of the way to a deeper 
insight into the underlying structures of string theory. It became clear 
that the five perturbatively distinct superstring theories are connected 
by a whole net of duality symmetries, whose higher--dimensional part is 
depicted in Fig.\ref{F-StringDualities}. With other words, it was shown 
that by compactifying any one of the five perturbatively distinct superstring 
theories on a suitable manifold and then de--compactifying it in another 
manner, one can reach any other of the five theories in a ``continuous'' 
way. The supposition arose that the five superstring theories in ten 
dimensions are just different limits of one unique and, thus, more fundamental 
theory. This theory was called by A.Sen {\it $U$--theory}, where the 
``$U$'' stands for ``Unified'' or ``Unknown'' \cite{Sen98b}. 

\begin{figure} 
\centerline{\epsfig{file=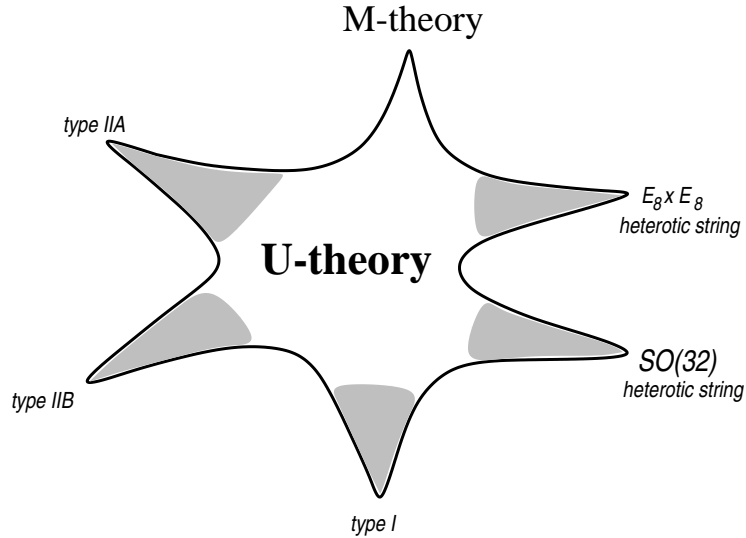,width=4in}}
\caption{\footnotesize 
The moduli space of $U$--theory, the unified string theory.
Here, the shaded regions denote the weak coupling limits 
in this moduli space and, thus, refer to the five perturbatively
distinct string theories. In the other regions such a description 
in terms of a weakly coupled theory --- where perturbation
theory is allowed --- is not possible. One of this corners is the 
limit where the coupling goes to infinity, which now is called 
$M$--theory.
\label{F-ModuliSpace}}
\end{figure}

In fact, the apparently different string theories and their compactifications 
can be viewed as just different limits in the parameter space (moduli space) 
of this central theory where the coupling is weak, as shown in 
Fig.\ref{F-ModuliSpace}. Here, the shaded regions correspond to weakly 
coupling limits in the moduli space, giving rise to the perturbatively 
different string theories. Moreover we see, that most of the regions in the 
moduli space of $U$--theory are not perturbatively accessible. One 
special limit, namely the limit where the coupling goes to infinity, is 
now known as $M$--theory. Here, to speak with the words of E.Witten, 
{\it `` `M' stands for `Magical', `Mystery' or `Membrane', according 
to taste''} --- or ``Mother'' (A.Sen). It was first introduced by E.Witten 
on a talk given at the University of Southern California in February 
1995.\footnote{At this point I want to note that $M$--theory is often 
identified with $U$--theory. However, here we shall keep in mind the 
distinction between the two: $M$--theory is a certain limit in the moduli 
space of the more fundamental $U$--theory, which also inherits the known 
weakly coupled string theories as certain limits.}
Its discovery during the investigation of the strong coupling limit of 
type $IIA$ strings was a real surprise \cite{Wit95a}. It turned out that in 
this limit certain non--perturbative objects, now called $D0$--branes, appear, 
forming a continuous spectrum and effectively generate an extra eleventh
dimension. This way, the type $IIA$ string theory at ultra--strong coupling 
gains eleven--dimensional Lorentz invariance. 

However, at present not very much is known about $M$--theory. As shown in 
Fig.\ref{F-StringDualities}, it arises in certain compactifications of type 
$E_8 \times E_8$ heterotic and type $IIA$ string theory. Moreover, the low 
energy limit of $M$--theory is given by the well--known eleven--dimensional 
$N=1$ supergravity, which in some sense can be viewed as the unique 
(although non--renormalizable) ``mother'' of all theories.
But due to the fact that $M$--theory is defined as that limit in 
the moduli space where the coupling reaches infinity ---
therefore it does not possess any longer a coupling constant or another 
free parameter --- up to now one does not have an appropriate mathematical 
description or, at least, a deeper physical understanding of this theory 
beyond the perturbative level given in the framework of its low--energy 
supergravity limit. First attempts to overcome this lack are 
stimulated by the observation that $M$--theory in a certain frame, namely 
the infinite momentum frame, is equivalent to a quantum mechanical system 
in the sense that scattering amplitudes in $M$--theory correspond to 
correlation functions in this quantum mechanical system. The fundamental 
degrees of freedom of this quantum mechanical system are given by 
$N \times N$ matrices, and its Hamiltonian is that of a certain 
supersymmetric quantum mechanics, where at the end of the calculation 
one has to take the limit $N \rightarrow \infty$ (see \cite{Duf98a} or 
\cite{Gib98a} for a short introduction).

\begin{figure} 
\centerline{\epsfig{file=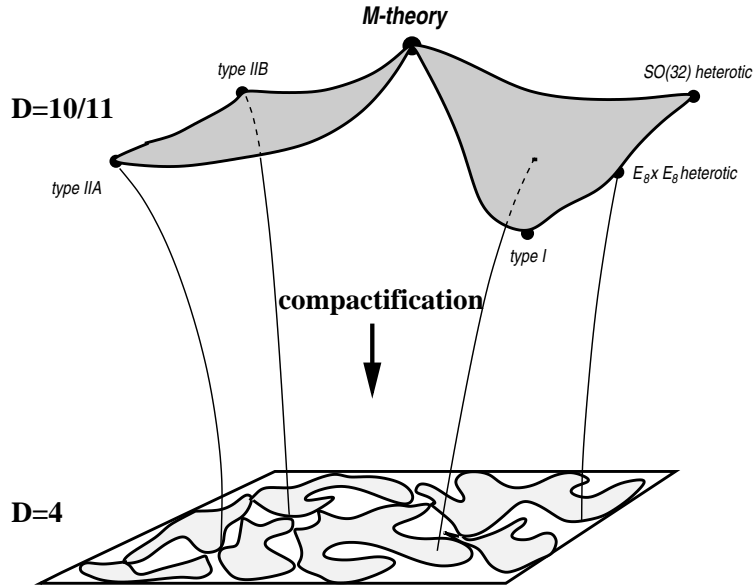,width=4in}}
\caption{\footnotesize 
The net of string dualities in higher dimensions manifests
also in lower dimensions after suitable compactifications.
Due to the plenty of possible compactifications, the resulting 
moduli space takes a much more complicated structure. 
In general, each region of the four--dimensional moduli space 
can be reached in several ``dual'' ways via compactification of the 
higher dimensional theories. Moreover, non--perturbatively all these 
vacua turn out to be connected (for instance by extremal transitions) 
and, this way, form a continuous web.
\label{F-ModuliSpaceLowerDimensions}}
\end{figure}

Although not fully understood, the main power of $M$--theory lies in 
its non--per\-tur\-ba\-tive description. Maybe no physicist doubts that 
perturbation theory is only and can only be a {\it small} step towards 
a more fundamental understanding of nature. $M$--theory itself is 
necessarily (by definition) a non--perturbative and, thus, an exact theory 
without any free parameters from which by duality transformations different 
regions in the moduli space of $U$--theory are accessible. This moduli space,
which describes the higher dimensional theories, is just a small piece of 
a much more extended moduli space obtained by compactifying the 
higher--dimensional theories on certain manifolds. Due to the huge 
amount of allowed ways to compactify down to lower dimensions, the 
resulting moduli space is significantly more complicated, as it is 
schematically depicted in Fig.\ref{F-ModuliSpaceLowerDimensions} 
\cite{Ler97a}.

However, even in lower dimensions one is faced with the lack of an exact 
mathematical description of string dualities, especially of non--perturbative 
symmetries. But one very important viewpoint also applies here: All theories 
in lower dimensions --- which can be obtained via compactification from the 
five ten--dimensional superstring theories --- seem to be connected by 
duality symmetries, either by continuous transformations, extremal transitions 
or the detour over higher dimensions. With other words, instead having to 
choose between many four--dimensional string theories (each one equipped with 
its own moduli space) --- which, among many others, also contain the standard 
model of strong, weak and electromagnetic forces as a possible solution --- 
we really have just one theory with, nevertheless, very many facets. 
Although this is only of little help to solve the vacuum degeneracy problem 
mentioned above (because it does not restrict the ways how to compactify), 
it gives for the first time in the history of physics (even the whole science)
a hint how a ``realistic'' unifying theory should look like: Each of the 
existing mathematical and physical models describing different aspects of 
our nature (and, thus, bearing some truth in it) can be viewed as just 
different aspects --- different realizations, if you want --- of one and 
only one underlying theory. This is similar to the concept of effective 
theories which can be obtained from a more fundamental exact theory. 
All those different effective models are adapted to describe distinct 
aspects within their limits. But instead of trying to unify the small 
number of realized or used effective theories, one can look for 
the underlying exact theory, which in the case of strings is given 
by the parameter--free $M$--theory. This indeed is a great discovery, 
not just from the viewpoint of the search for a real unifying theory 
(or TOE). The concept of duality symmetries opens up a complete new way of 
thinking about strings, unifying theories and the physical description of 
the nature as a whole. 

But this new way of looking at physical theories (namely string theories in 
various dimensions) in some sense as just different realizations of one and 
the same exact theory (adapted to the physical phenomena they intended to 
describe and only valid within certain limits) is only one part of the 
story. The concept of duality and the search for proofs of duality conjectures
has led to completely new developments also in another sense: Some formerly 
rather sharp separated and more or less independent investigated aspects of 
(mathematical) physics --- like classical and quantum properties of field 
theories, solitonic solutions of classical field equations and the fundamental 
degrees of freedom of quantum theories, singular classical objects (for 
instance black holes) and new types of topological defects, namely $D$--branes 
and $p$--branes --- are now starting to appear in a new and unexpected 
unifying light.

For instance, as mentioned above, duality often relates a weakly coupled 
(string) theory to a strongly coupled (string) theory. This way a perturbative 
expansion in one theory contains information about non--perturbative effects 
in the dual theory and vice versa, making duality a property of the full 
quantum theory and not just of its classical limit. However, because quantum 
objects in one theory get mapped via the duality transformation to rather 
classical solutions, like solitons, the distinction between classical and 
quantum objects loses its significance. At the same time, under a certain 
duality map, an elementary particle in one theory can be transformed into a 
composite particle in the dual theory and vice versa, making the strict 
classification of particles into fundamental and composite ones less 
meaningful.

\begin{figure} 
\centerline{\epsfig{file=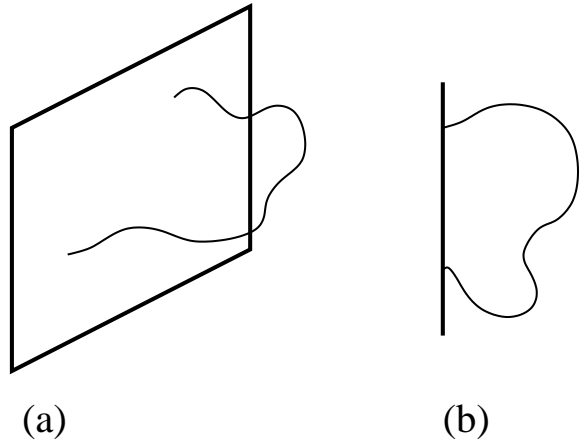,width=3in}}
\caption{\footnotesize 
Open string states with ends attached to a $D2$--brane 
(Dirichlet--membrane) and a $D1$--brane (Dirichlet string).
\label{F-DBranes}}
\end{figure}

Another way of development is given by the explicit occurrence of higher 
dimensional objects within the framework of string theory. A special type of 
solutions of the string field equations are $p$--dimensional objects, called 
$p$--branes, whose quantum dynamics can be described by a $(p+1)$--dimensional 
QFT. Here a $p$--brane denotes a static configuration which extends 
along $p$ spatial directions (tangential directions) and is localized in all 
other spatial directions (transverse directions). Thus, a $0$--brane is a 
point--like object (particle) which sweeps out a one--dimensional world line. 
Analogously, strings are $1$--branes sweeping out a two--dimensional 
world--sheet, membranes are $2$--branes and so on. 

A special type of $p$--branes, namely topological defects on which the 
ends of a string can be trapped, are called Dirichlet $p$--branes or, 
simply, $D$--branes (see for instance \cite{Pol96b}). They were first 
discovered in the study of perturbative dualities of string theories and 
manifest itself as extended solitonic objects. In the presence of these 
solitons there can be open string states whose ends lie on these extended 
objects, see Fig.\ref{F-DBranes}. This way, the open string dynamics can 
describe the internal dynamics of the $D$--brane to which it is attached, 
yielding a quantum field theory of higher dimensional objects. Such field 
theories were formerly considered to be inconsistent and suffered from deep 
anomalies. Indeed, it is now known that in addition to the fundamental 
strings, non--perturbative string theory must contain a rich spectrum of 
branes in order to be consistent. Thus, also the strings lose their 
significance as the fundamental objects of string theory; one has
to describe the underlying theory using in addition a plenty of other
objects. This is without doubts a further step towards a real unification!
String theory was just our entrance to that new and fascinating description of 
nature where different aspects magically get unified. 

This unifying concept of formerly nearly unrelated fields of theoretical 
physics becomes most obvious in the case of the black holes of general 
relativity and the solution of the black hole information paradox within the 
framework of string theory. Black holes are long known as singular solutions of 
the classical field equations of gravity. In the early 1970s it was found that 
black holes also obey laws analogous to the laws of thermodynamics. This opened
up the way to a ``quantum theory'' of black holes, which was initiated by the 
famous discovery of the Hawking radiation in 1975: Black holes radiate as 
black bodies at the corresponding temperature. In addition, the entropy of 
a black hole was found to be given by the Bekenstein--Hawking formula. 
However, until recently there was no known way to count the states of a 
black hole to give a microscopic interpretation of this entropy (as it
is known from conventional statistical mechanics). Moreover, due to the 
absence of such a microscopic description in the case of thermal radiation 
from a black hole, one is faced with the so--called {\it black hole 
information paradox}: A black hole of a definite mass and charge can be 
formed in a very large number of ways. The final state after its evaporation 
is given by black body radiation and does not depend on how the black hole 
was formed. That is, many initial states evolve into a single final state.
But this violates the known laws of quantum mechanics and thermodynamics!

\begin{figure} 
\centerline{\epsfig{file=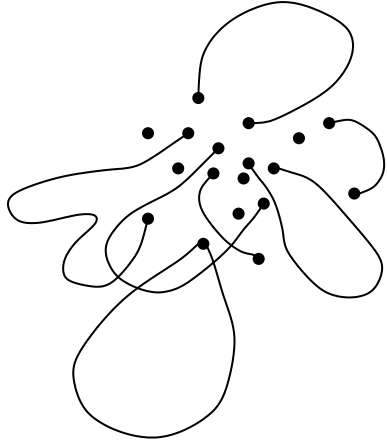,width=2in}}
\caption{\footnotesize 
A state of many $D0$--branes with strings attached.
When the coupling becomes strong, the $D0$--branes get modified 
by gravitational self--interactions in such a way that a classical 
black hole space--time becomes the right description.
\label{F-BlackHole}}
\end{figure}

A lot of solutions to this problem were proposed (for a short summary see 
\cite{Rov98a}), but only recently a solution was presented which does not 
lead to a change of the basic laws of physics (such as the laws of quantum 
mechanics or the locality principle in QFT) and to the introduction 
of mystical states remaining after the evaporation of the black hole. 
This solution comes within the framework of string theory: For a special type of 
charged black holes, so--called extremal black holes, the counting of 
microscopic states, namely BPS states, were carried out. It was shown that 
this black holes have a perturbative description in terms of a collection 
of $D$--branes at weak coupling. When the coupling becomes strong and, thus,
this perturbative description breaks down, the $D$--branes become modified 
by gravitational self--interactions in such a way that a classical black hole 
space--time appears. At weak coupling one uses perturbation theory to count 
the number of quantum states of the strings propagating along the 
``surfaces'' of the $D$--branes. Due to the fact that the degeneracy 
of BPS states does not change when going to strong couplings, that number 
must also be valid for strong coupling, i.e. the black hole space--time. 
As a surprise it turned out that the answer one finds corresponds exactly 
to the Bekenstein--Hawking entropy of the black hole! Thus, string theory 
provides a ``natural'' microscopic explanation of the black hole entropy 
without giving up too much fundamental concepts of physics. Moreover, one 
can compute the rate of Hawking radiation from these black holes due to 
quantum scattering processes inside the hole. Even here this rate agrees 
with the Hawking radiation. 

Although the calculations are carried out only for the specific type of 
extremal black holes one hopes that this arguments can also be applied 
to the general case. No matter how the results will look like, this very
concrete application of string theory shows that it is more than just an 
idea thought up by some ``crazy'' theorists. String theory opens up the 
possibility to understand up--to--now unsolved problems in theoretical 
physics, and without doubts it sheds more light into the darkness 
surrounding the way towards a more fundamental understanding of nature.

\section{Resume or {\it What have we learned so far?}}
\label{Resume}

This brings me straight to the question of what the conclusion of all this 
might be. Of course, currently no one can give a satisfying answer to this 
question because the analysis of that new and amazing discovery called 
string theory just started. But some general aspects and concepts became 
already clear:
\begin{itemize}
\item String theory, originally regarded as an ultraviolet finite
      way to unify classical gravity with quantum mechanics, now became 
      a --- or {\it the} (at least from the viewpoint of a string 
      theorist) --- leading candidate for an unifying theory of the 
      known fundamental forces of nature. Within its framework, the 
      concept of duality symmetries arose to one of the main physical 
      principles (not only of string theory, but of nature at all),
      showing that the five perturbatively different superstring theories 
      in ten dimensions (which are the result of the first superstring 
      revolution in the 1980s) can be unified and are just different facets 
      of one unique underlying theory, called $U$--theory. 
\item In the moduli space of $U$--theory there appears in the limit of 
      infinite coupling a new eleven--dimensional theory, called $M$--theory, 
      with no free parameters or couplings. In the mathematical and physical 
      understanding of this exact theory, whose description can only be 
      achieved non--perturbatively, might lie the key to the TOE.
\item Duality symmetries are not restricted to higher dimensions but also
      appear after compactification in lower dimensions, even in the real 
      four--dimensional space--time. They lead to remarkable symmetries 
      between formerly unrelated physical theories and models. As in the 
      case of string theories in higher dimensions, duality shows that 
      lower--dimensional effective theories are just different limits in 
      the moduli space of a unique theory, adapted to describe certain 
      physical aspects of nature. 
\item The investigation of dualities opened up completely new ways between 
      formerly rather unrelated or strictly distinguished aspects of 
      theoretical physics (classical and quantum, composite and elementary, 
      smooth and singular, strings and $p$--branes). 
\item Non--perturbative dualities take us beyond string theory and the 
      perturbative description of nature.
\end{itemize}

To summarize, string theory has proven to be {\it a unifying theory} allowing 
a deeper understanding of the leading physical principles of nature. Moreover, 
the developments of the last years have shown that it provides powerful tools 
whose application will bring us beyond the current description of our world 
within the framework of physical theories. Thus, without doubts, string theory 
opens up a possible new way towards one of the great goals of theoretical 
physics, namely the formulation of {\it the unifying theory} --- or TOE. 
Even if that hope will never come true in the future, string theory has shed
and will shed light into many formerly dark areas of theoretical physics.
So, it cannot be completely wrong! 

String theory is  --- at least compared with the ordinary quantum field
theoretical researches --- a very young development; we all (especially those 
who do not believe in that ``crazy'' string theory) should give it some
time to prove itself useful \ldots


\end{document}